# A Tree-based Approach for Detecting Redundant Business Rules in very Large Financial Datasets

**NhienAn LeKhac, Sammer Markos, M-Tahar Kechadi**
*University College Dublin, Ireland*

**ABSTRACT**
Net Asset Value (NAV) calculation and validation is the principle task of a fund administrator. If the NAV of a fund is calculated incorrectly then there is huge impact on the fund administrator; such as monetary compensation, reputational loss, or loss of business. In general, these companies use the same methodology to calculate the NAV of a fund, however the type of fund in question dictates the set of business rules used to validate this. Today, most Fund Administrators depend heavily on human resources due to the lack of an automated standardized solutions, however due to economic climate and the need for efficiency and costs reduction many banks are now looking for an automated solution with minimal human interaction; i.e., straight through processing (STP). Within the scope of a collaboration project that focuses on building an optimal solution for NAV validation, in this paper, we will present a new approach for detecting correlated business rules. We also show how we evaluate this approach using real-world financial data.

*Keywords:* Net Asset Value; NAV validation; business rule validation; tree topology; correlated rules, straight through processing.

**INTRODUCTION**
Fund administration is a subsection of the investments industry – i.e., trading on the stock exchange (Abner D. J. 2010). When it comes to trading on the stock exchange there are three tiers administration, at the front line is front office- these are the traders, they trade on behalf of their clients. The second tier is middle office, where all the actual trade information and other data are tabulated and sent in an agreed format to the final tier; back office (Gross S. 2006). Back office is the accountancy work that is carried out for all trades as per front office's instructions; hence the price or Net Asset Value (NAV) of a fund is calculated. There are many steps in the pricing or valuation of a fund. This depends on what type of fund it is, however, all funds consist of six core pieces: (1) Stock reconciliation, (2) reflection of corporate actions, (3) pricing of instruments, (4) Booking, calculating and reconciling fees and interest accruals, (5) cash reconciliation and finally (6) NAV/price validation.

The NAV validation (CFA Institute. 2008) price is as vital as this is the verification for accuracy and correctness occurs. If a NAV price released to a client is incorrect then the fund administrator is liable according to the fund administration agreement. In this case there may be hundreds of thousands of monetary compensation to investors and the clients.

Besides, most fund administration companies in Ireland are located in the financial district IFSC (International Financial Services Centre-Dublin); therefore are regulated by the financial regulator (FR) and are audited yearly. So according to the Irish and EU directives there are many checks and standard reports that must be available, these reports are always checked for manual intervention. As this is an international fund administration company, their operations are spread throughout the world; i.e., Europe, Asia, North America, etc. Therefore there is an urgent need for an improved approach to maintain and track all operations across this distributed environment. Therefore, the fund administration industry is trying to reduce the operational cost of NAV validation process.

In the NAV validation step (step (6), cf. page 1), results from previous steps in producing a NAV (from (1) to (5), cf. page 1) are checked. The data that are reflected on the fund accounting system are checked against broker statements, pricing vendor reports, cash statements, etc. This step is normally completed by eye balling the reports and making sure that the external reports match what is reflected on the NAV (in the fund accounting system). If the data does not match or is not within the allowable tolerance dictated by the prospectus then valid evidence must be attained to answer why not, this evidence is attached as a hard copy within the file for audit purposes. Each fund has its own set of business rules attached to it. The selection of rules is based on the attributes of a fund.

Recently, in the context of reducing operational cost, the fund administration business has raised some questions on the optimisation of rule sets. Concretely, how can the user optimise their performance, minimise their running expense by selecting a reduced set of rules without compromising the NAV validation. This analogy leads to many challenges. We firstly need to have a strong knowledge of the given business. Secondly, there is a need to analyse the relationships among these rules and these relationships are normally not straightforward. This step requires a fund accountant to verify each rule against the total population of rules applied to a given fund. In our recent study, it took an expert one working day (8.5hours) to analyse the correlation within a set of 20 rules. It takes then approximately a further 4 working days to verify these results using real-world data retrieved from a NAV validation system. In reality, in order to carry out the NAV validation of a fund, a set of 50 to 100 business rules is applied. Note that the complexity of this task is $O(n^2)$ where $n$ is the number of rules. Moreover, a fund administration business can run thousands of funds in multiple regions around the world. In addition, new funds launch daily and hence a set of rules needs to be attached to this new entity. As a consequence, the new set of rules needs to be scrutinised.

In this paper, as part of collaboration between our research laboratory and an international investment bank within the context of creating a BI-Based organisation (Wixom B., Watson H. 2010), we propose a tree-based solution for detecting correlated rules within a rule set that can be applied to a fund. This solution is built as a software tool that can assist users for automatically validating their rule sets. We also evaluate our approach using real data to illustrate its efficiency.

The rest of the paper is organised as follows. In Section 2 we discuss the background research work that is related to our approach. Section 3 presents the purpose of the business rule validation as well as its issues. We describe in detail our tree-based solution in Section 4. Section 5 we evaluate the proposed approach on real-world data obtained from BEP bank. In Section 6 we discuss perspectives and conclude.

# BACKGROUND

In this section we present firstly the process of NAV validation in financial institutions. Secondly, we briefly describe the concepts related to business rules. Finally, we discuss the use of tree topology to index data.

## NAV validation and preparation in banking and finance

NAV is a price for an entity that is published on the stock exchange, which then allows investors to buy shares; invest in this entity. Fund administration is the name given to the set of activities that are carried out in support to the process of running a collective investment scheme, whether the scheme is a traditional mutual fund, a hedge fund, pension fund, unit trust or something in between. In this study, these administrative activities would include the calculation of a fund NAV (investment scheme or pool of money). Some investment management companies calculate their own NAV. However, if it is an Irish regulated fund then its NAV must be calculated and verified independently (EU UCITS Regulations 1989). NAV can be calculated daily, weekly, bimonthly, quarterly, and semi-annually. Therefore, the NAV frequency is dependent on the characteristics of the fund Gross S. (2006).

The core steps in calculating NAV on a specific day are: stock or portfolio reconciliation; executing any potential corporate actions on some of the securities held on the portfolio, pricing the securities/instruments on the portfolio, booking, calculating and reconciling fees and interest accruals. These could be legal fees, audit fees and some NAV based fees (calculated on the size of NAV), such as administration fees, management fees, and performance fees (the last two are more common in hedge funds rather than in mutual funds), reconciling the cash accounts of the fund and finally validating the NAV.

**Stock or portfolio reconciliation.** This is the first step in NAV preparation; this is where the fund accountant will collate and reflect on all shares bought and sold by the investment/portfolio manager. This step is either completely manual or has a considerable amount of manual intervention.

**Execution of any potential corporate actions on some of the securities held on the portfolio**. Once all trades/shares are reflected, corporate actions such as acquisitions, mergers, ISIN changes, dividends, etc., must be also reflected on the fund. These can directly impact on the funds cash (i.e., dividends payable/receivable), and also the number of holdings/name/entity (i.e., ISIN change, merger, etc.). This is also completely manual process; no automation at all, hence prone to human error.

**Pricing the securities/instruments on the portfolio**. Once all stock positions and corporate actions are reflected on the fund accounting system, the next step is to price the stock/securities held on the fund portfolio. Pricing is a very important issue to many back office fund administrations because they must be independent of the prices. They must reflect the price according to the legal contract or prospectus. This legal document outlines the strategy of the fund and that includes pricing sources and price types. For example, the pricing hierarchy of equity in fund X is bid price source Extel (http://www.extelsurveys.com/), Telekurs (http://www.six-telekurs.com/tkfich_index/tkfich_home.htm) then Bloomberg (http://www.bloomberg.com). This means that the fund administrator must first try to source the price from Extel if there is no price available from this source then they try Telekurs and so on. This is a semi-automatic process.

**Booking, calculating and reconciling fees**. Some of the fees are legal fees, audit fees and some NAV based fees (calculated on the size of NAV), such as administration fees, management fees and performance fees. On the income side, these fees are more like credit interest accruals and they are split into different parts (i.e. different fees) when it comes to fees. First, calculate the correct amount of the fund at a particular NAV date. For instance, for a monthly NAV, the audit fees are €12,000 per annum, so the amount that should be reflected as an accrual on the NAV is €1,000, however if these fees are prepaid then the amount that would be reflected is a prepaid expense that is reduced every month. Another scenario is that if the fund pays the investment manager 1,000 of the 12,000, then this amount must be reflected as leaving the fund cash account and a reduction in the accrual (more details are given in cash reconciliation section below). This process is semi-automatic, the accrual is calculated by the accounting system, however, any payments are reflected manually.

**Reconciling the cash accounts of the fund**. This is the final stage of the NAV preparation and it deals with reconciling all the cash accounts that the fund holds. These accounts can be multicurrency and be held in multiple locations – for instance a EURO account held in JPMorgan an a US$ margin account held in Goldman Sacs. Before the recession that hit the economy the funds used one broker, now the common investment manager strategy is to spread their liquid cash across many brokers to reduce the exposure to high risk. This step is reconciling all cash statements with what is reflected on the fund accounting system.

**Validating and publishing NAV**. Once all the core operations of the NAV are executed, we proceed to the validation process of the NAV. This is the most time consuming and at present it is purely a manual process. In this step, all points carried out in producing a NAV are checked. Currently, there is too much emphasis on manual process and, therefore, there is a need for human resources as it is labour intensive and also an automated solution will reduce the risk of errors.

## Fund Rules

**Rule**. A rule is a knowledge representation technique and its structure relates one or more conditions, antecedents or a situation to one or more conclusions (consequents) or actions (Gaˇsevi´c D., Djuri´c D., Deveďzi V. 2006). The premises are contained in the IF condition of the rule, and the conclusions are contained in the THEN statement, so that the conclusions may be inferred from the premises when they are true. For instance, IF *time is 1pm* THEN *we break for lunch.* Premises are typically represented as facts, Actions can assert a new fact or can perform an operation (e.g., invoke a procedure).

**Business Rule**. A business rule (Amghar Y., Meziane M., Flory A. 2000) is a statement that defines or constrains some aspects of the business (Von Halle, B. 2001). It is intended to assert business structure or to control or influence the behaviour of the business. Business rules describe the operations, definitions and constraints that apply to an organisation. Business rules can apply to people, processes, corporate behaviour and computing systems in an organisation. They are put in place to help the organisation achieve its goals. For example, a business rule might state that no credit check is to be performed on return customers, a rental agent can not allow a rental tenant if their credit rating is too low, or company agents can use a list of preferred suppliers and supply schedules, etc.

**Tree Topology**

Tree topology is one of the most favourable methods for representing and indexed data, as the search operations are linked with tree nodes. The tree topology can be of any degree; from a binary (Knuth D. 1997) to a B-Tree family such as B/B*/ B+Tree (Bayer, R. 1971, Comer D. 1979, Hans B. 1979), S-Tree (Tousidou E., Vassilakopoulos M., Manolopoulos Y. 2000, Shvachko. K.V. 2004). For instance, some DBMSs implement an index structure based on B-Tree such as MySQL (Colin C. 2008), SQL Server (Raghu R., Johannes G. 2000). Nevertheless, this topology is not efficient enough for indexing complex, heterogeneous, and noisy data. However, tree topology can be used to parse mathematical expressions efficiently.

## BUSINESS RULE VALIDATION

**Objectives**: The main objective of business rule validation is to analyse the rules for correlation, duplication, and redundancy. This is to ensure optimal efficiency of NAV before releasing it to the clients. Rule validation is made up of two phases: 1) examine the validity of the rules in context to a superfund; 2) investigate the possibility of relationships between rules (association/correlation). For example we can have a scenario where rule 1 and rule 2 always occur at the same time and weather this is a duplicate or not.

**Challenges**: There are many challenges in validating of business rules, such as large number of rules, the analysis of the content of each rule, which is usually very complex and difficult to extract the right attributes. But one of the difficulties at the moment is that there is no existing solution in the market in order to allow a comparative study and benchmarking.

## METHODOLOGY

**Business rule representation**

There are different methods for representing business rules. We can represent a business rule by describing its content. For instance, rule HLD01 is *"Will fail the test if the cost (local or base) of the holding is zero"*. This method is simple, however it is difficult to automatically validate because it conceals crucial information; e.g., how we can calculate the *cost of the holding*? Therefore additional information from fund accountants is needed in this case.

Business rules can also be expressed in formal languages such as Unified Modelling Language, Z notation, Business Process Execution Language, Business Process Modelling notation, etc. However, all these formal languages aim to represent not only business rules but also the business process in which these rules are applied. As a consequence, they add unnecessary complexity to the business rule representation.

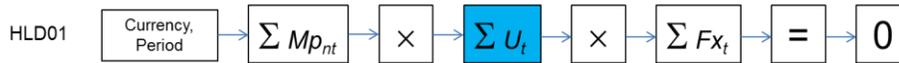

*Figure 1. Linked List representation of HLD01*

In our approach we are looking for a simple but efficient representation of a business rule; hence we apply the "IF… THEN …" structure. As mentioned in Section 2.2, this structure relates one or more conditions to one or more conclusions. For instance, rule HLD001 can be represented as "IF *cost of the holding is zero* THEN *fail*". Furthermore we normalise the condition of this structure (using some mathematical expressions). For example, rule HLD001 can be expressed by "IF $C_{nt} = 0$ THEN *Fail*" with $C_{nt}$ – cost of the holding, $C_{nt} = \Sigma( TX_{nt}$ x $U_t$ x $Fx_t)$ , $TX_{nt}$ – Transaction price, $U_t$ – unit/# of shares, $Fx_t$ – exchange rate. This means that the first step of the pre-processing phase is to normalise the condition of the rules by using a mathematical expression with pre-defined parameters. This step can be done manually or automatically. In this paper, a manual method is considered with the help of a fund's expert who can analyse the content of the rules. We combine this structure with the context of the business rule in order to represent it completely. This context describes different factors that affect the execution of this business rule in relation to validating the NAV of a fund. For example, time period and foreign currency are applied to rule HLD001.

We then apply an appropriate data structure to represent the rules. This allows us to validate them more efficient than in the "IF… THEN…" format. The first data structure considered is a linked list. As shown in Figure 1, this linked list can represent both the content and context of a rule. However, the linked list cannot clearly reflect the relationships between the rules. Therefore we propose a tree-based method to represent the rules. We describe this method in the next section.

**Tree-based Representation**

The general format of a tree-based representation has three main parts: 1) the body of the business rule; 2) the context or domain of application of the rule, and 3) the value used to verify the rule (Figure 2).

We use a grammar, defined in Figure 3, to build the body of a business rule. The body is represented by a mathematical expression that includes two elements: operands and operators. Figure 3 also shows the definition of each element as well as how to generate an expression. Figure 4 defines the grammar for the context.

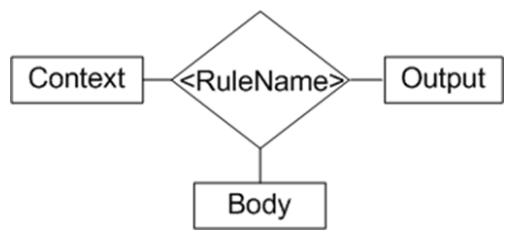

*Figure 2. Tree-based representation of a business rule.*

```
Expr := <Operands> | <Operators>
<Operators> := + | - | × | ÷ | Σ*
<Operands> := <Expr.> | <Parameters> | <Values>

<Parameters> := exchange rate, trade date, market price, etc.
<Values> := 365, etc.
```

Figure 3. Grammar of the tree body.

```
Context := <Output> and/or <Logical expression>
<Output> := <Relational operator> <Values>
<Logical expression> := <Operands> | <Operators>
<Operators> := ∩
<Operands> := <Logical expression> | <n> | <t_{j-1}> | <t_j> | "bond" | "equity" | "future"

<Relational operator>: = | < | > | ≥ | ≤ | ≠
<Values> := 0, 0.2, etc.
<n> := local ccy, base ccy, etc.
<t_j> := start date
<t_{j-1}> := end date, t_{j-1} < t_j
```

Figure 4. Grammar of the tree context.

## Working Example

This section presents examples of tree-based representation based on the technique described in the previous section. Figure 5 to 7 are tree-based representations of rules from HLD01 to HLD04 respectively.

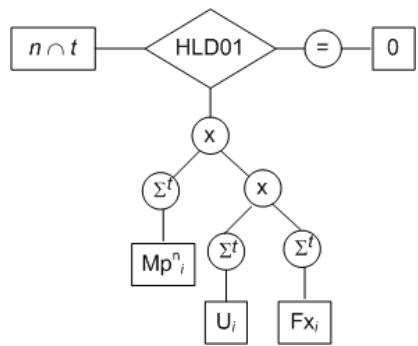

Figure 5. Tree-based representation of rule HLD001.

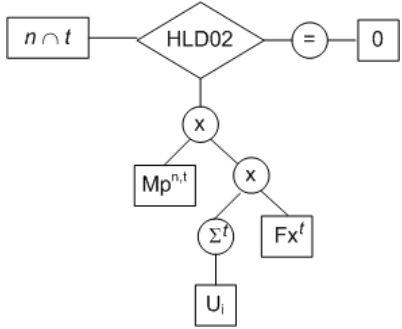

*Figure 6. Tree-based representation of rule HLD002.*

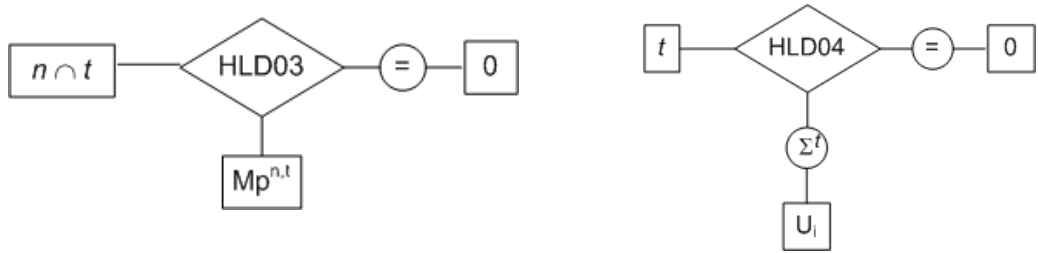

*Figure 7. Tree-based representation of rule HLD003 and rule HLD004.*

### Algorithm

In this section, we describe a tree-pruning algorithm that is used to detect the relationships among business rules. The main goals of this algorithm are: detecting the relationships among different rules in a rule set, building the map of rules based on their relationships and validating the rules to create the optimal rule set as well as the correlated rule set.

Before describing the algorithm, let's formally define some concepts:

**Definition 4.1** Given a rule set $P$, $\forall$rule $R_i, R_j \in P$. $T(R_i)$ is the tree representation for the body of rule $R_i$ and $h_T(R_i)$ is the height of a tree $T(R_i)$. There is a relationship between two rules $R_i, R_j$ if

- $T(R_i) \subset T(R_j)$
- $T(R_j) \subset T(R_i)$

**Definition 4.2** Given a rule set $P$, $\forall$rule $R_i, R_j \in P$. A rule $R_i$ is called a core rule if

- $h_T(R_i) = 1$
- If $h_T(R_i) > 1$ then $\forall R_j \in P: R_j \neq R_i$, $T(R_j) \not\subset T(R_i)$

A rule set C is called the core rule set if $C \subseteq P$; $\forall R_i \in C$, $R_i$ *is a core rule*. A rule $R_j$ is called a dependant rule $R_j$: $\exists R_i \in C$, $T(R_i) \subset T(R_j)$. A rule set $D$ is called a dependant rule set if $D \subseteq P$: $\forall R_i \in D$, $R_i$ is a dependant rule.

The tree-pruning algorithm is defined in the following:

**Algorithm 4.1**
*Input*: Rule Set P
*Output*: Core Set C, Correlated set D
*Step 1*. $C = \emptyset, D = \emptyset$
*Step 2*. For each rule $R_i \in P$
      If $R_i$ is a core rule (*wrt.* Definition 4.2) then
          $\{C\} \leftarrow \{C\} \cup R_i$
      Else   $\{D\} \leftarrow \{D\} \cup R_i$
      EndIf
    End For
*Step 3*. For each rule $R_i \in C$
      If *Value_Check($R_i$, $\{C\} \setminus R_i$)* then
          $\{D\} \leftarrow \{D\} \cup R_i$
      EndIf
    End For

   The function Value_Check(Ri , {C}\ Ri) is used to verify whether rule Ri fails or not when any rule in {C}\ Ri fails in terms of the value of these rules. Figures 8 to 10 show some examples of the tree-pruning algorithm after Step 2.

## EVALUATION

We implemented the Algorithm 4.1 in a system named ANT, this was the environment used to validate the business rule sets. In this section we evaluate our approach with various rule sets. The first rule set is selected from the fund K7.

## Experiments

The first rule set used in the NAV validation process for the fund K7 has 47 rules. The names of these rules are: HLD12, HLD13, TXN08, TXN18, CFL103, CFL104, HLD104, TXN101, ACC001, ACC106, HLD102, HLD103, HLD109, HLD112, HLD113, HLD114, PR115, PR116, PST100, TXN102, TXN103, TXN109, TXN110, TXN111, TXN112, TXN113, TXN114, TXN115, TXN116, TXN117, TXN118, TXN119, TXN120, TXN121, TXN122, TXN123, TXN125, TXN126, TXN127, TXN129, TXN130, HLD01, HLD02, HLD03, HLD04, HLD05, HLD09. Figure 11 shows the core set and the correlated set given by running the ANT system. By observing this Figure, we notice that there are 8 rules (19%) in the correlated set and 39 rules in the core set (81%). The names of correlated rules are TXN119, TXN120, TXN123, TXN124, HLD009, HLD113, HLD001, and HLD002. The second rule set used in this experiment consists of 145 rules. We notice that there are 72 core rules versus 73 correlated ones.

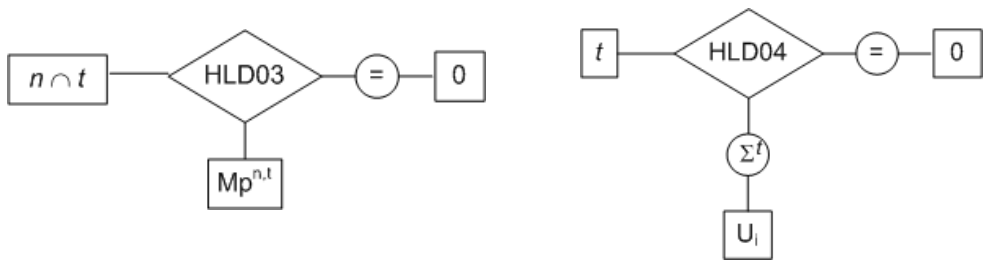

*Figure 8. Core rules: HLD003 and HLD004.*

The ANT system also has a visualisation tool that shows the tree representation of each rule as well as the relationship among the rules, so that the users can verify the output sets. For instance, Figure 12 and Figure 13 show the tree representation of rules HLD012 and HLD002 respectively. By observing these figures, we notice that both rules HLD012 and HLD002 are correlated to rule HLD003. Figure 14 shows the relationship between the rule HLD003 and its correlated set of rules. It also confirms that if rule HLD003 is used, we do not need to apply two rules HLD002 and HLD012.

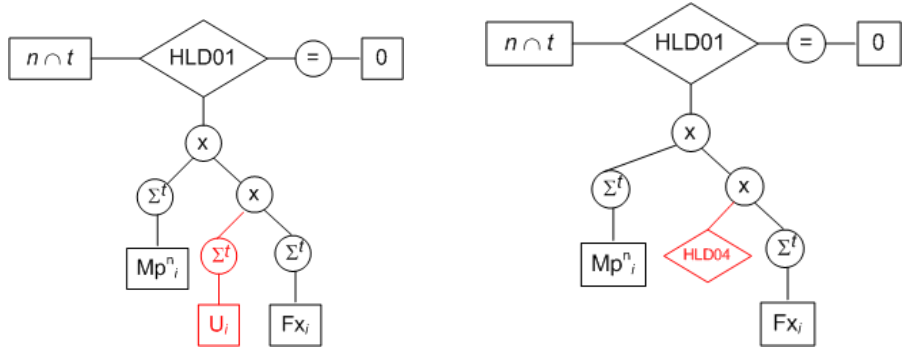

*Figure 9. Tree pruning algorithm on HLD001.*

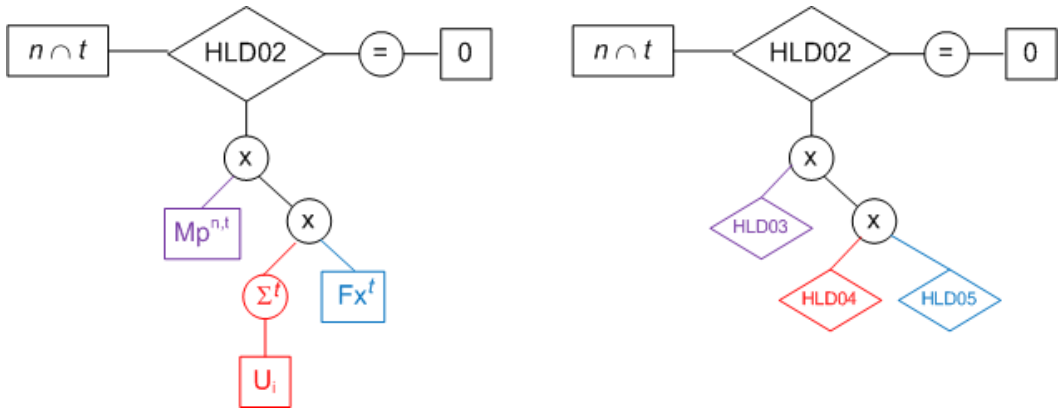

*Figure 10. Tree pruning algorithm on HLD002.*

In order to verify the output of the ANT system, we perform statistical analysis and confirmed that HLD002, HLD003 and HLD012 are correlated.

### Analysis
The tree-based solution proposed in this section has the following advantages:
- It represents rules in a simple and intuitive way.
- It can efficiently detect and verify relationships amongst rules.
- It can also determine sets of correlated rules.

Finally, this approach can help the users to analyse the rules for correlation, duplication, and redundancy.

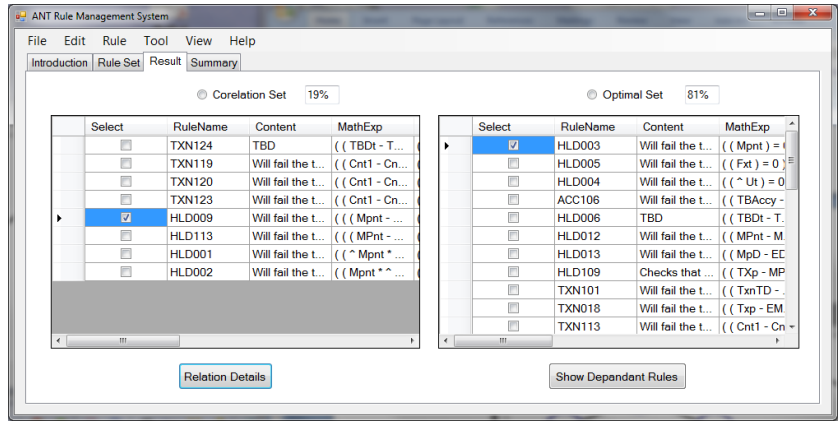

*Figure 11. Results of ANT system for K7.*

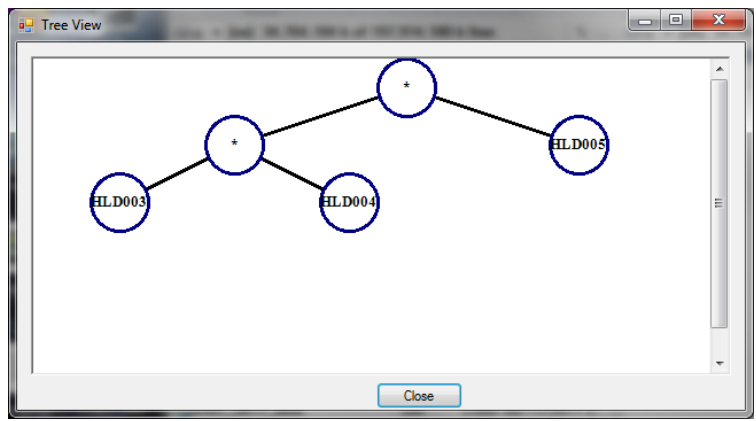

*Figure 12. Tree representation of Rule HLD002 by ANT system.*

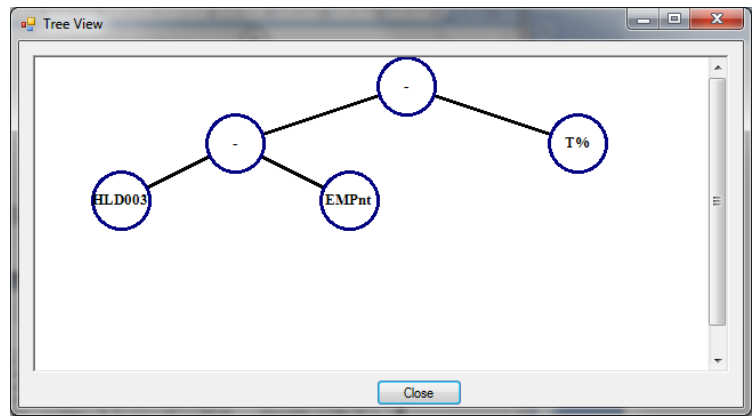

*Figure 13. Tree representation of Rule HLD009 by ANT system.*

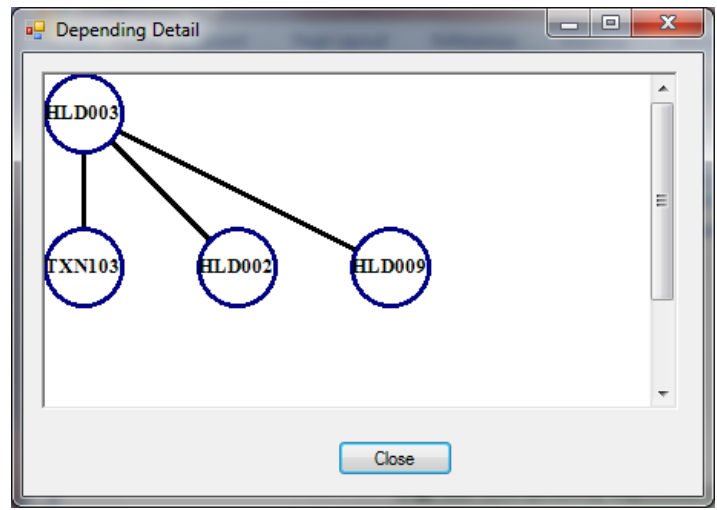

*Figure 14. Relationship between Rule HLD003 and its correlation by ANT system.*

## CONCLUSION AND PERSPECTIVES

In this paper we presented a new approach for detecting redundant business rules that can be found in the process of NAV validation. By using tree-based representation, we can not only intuitively represent these rules (already mapped to mathematical expressions) but also detect the relationships amongst them. Based on these relationships we can build groups of redundant rules. By using real-world data from financial institutions for testing, we show that this approach is efficient, robust and easy to use. However, the performance of this solution is totally dependent on the quality of mathematical expressions of the rules. It is worth noting that the original rules are expressed as text, converting the text to mathematical expressions is not a straightforward task. As part of pre-processing phase, this conversion is required in order to use the rest of the system.

This research and development has opened many perspectives:
- The current approach needs a manual standardisation where business rules expressed in text format are converted to mathematical expressions. Fund accountant experts perform manually this conversion and it is a very time-consuming task. For instance, it takes more than a month to map 100 business rules in our last experience. Moreover, this process is repeated when we need to validate different investment funds where different rule sets are applied. Therefore, we need an automatic or semi-automatic solution for this problem. Although all business rules are in text format, they have the same format. For instance, all rules start with "Will fail if…". Furthermore, all business rules are related to predefined business concepts such as holding cost, market prices, market value, exchange rate, NAV, etc. and their evaluated values. Hence, we can apply intelligent-based text processing techniques (Feldman R., Sanger J. 2007) as well as data mining techniques (Lawrence K.D, Pai D.R., Lawrence S. M. 2010) to analyse these rules and convert them to the mathematical expression mode automatically. This solution can improve the normalisation step of our approach.
- We also consider a scalability issues for the rule validation in terms of enterprise intelligent (Morabito J., Stohr E.A, Genc Y. 2011). Let's consider a scenario where this process will be applied in hundreds of branches of the investment bank in different countries across the world. Each branch could manage hundreds of investment funds with its own rule set. In this case, our solution should take into account the distributed and heterogeneity aspects the rule sets. An intermediate layer is required to negotiate as well as to integrate the rules into a homogenous set of rules and we can then verify and validate them.